\documentclass{PHYEAUTH}
\usepackage{graphicx}
\usepackage[english]{babel}
\usepackage{amsmath}
\usepackage{amssymb}
\usepackage{amsmath,amsfonts,bm,verbatim}

\newcommand{\rlangle}{\rangle\!\langle}

\begin{document}

\begin{frontmatter}

\title{Non-Markovian signatures in the current noise of a charge qubit}

\author[address1]{Alessandro Braggio\thanksref{thank1}},
\author[address2]{Christian Flindt},
and
\author[address3]{Tom\'a\v s Novotn\'y}

\address[address1]{LAMIA-INFM-CNR, Dipartimento di Fisica, Universit\`a di Genova, Via Dodecaneso 33, I-16146, Genova, Italy}

\address[address2]{MIC -- Department of Micro and Nanotechnology,
             NanoDTU, Technical University of Denmark, Building 345 East,
             DK-2800 Kongens Lyngby, Denmark}

\address[address3]{Department of Condensed Matter Physics,
     Faculty of Mathematics and Physics, Charles University,
     Ke Karlovu 5, 121 16 Prague, Czech Republic}

\thanks[thank1]{Corresponding author. E-mail: braggio@fisica.unige.it}

\begin{abstract}
We investigate the current noise of a charge qubit coupled to a
phonon bath in different parameter regimes. We find, using the
theory of Full Counting Statistics of non-Markovian systems, that
the current fluctuations are strongly influenced by memory effects
generated from the interplay between quantum coherence and the
dynamics of the phonon bath.
\end{abstract}

\end{frontmatter}

The recent experimental realizations of charge qubits in double
quantum dots clearly illustrate the importance of controlling
quantum coherence within a system \cite{Fujisawa:1998}. Detecting
quantum coherence via indirect and non-invasive measurements is thus
becoming a crucial task. Currents fluctuations, in particular the
current noise, have been suggested as a useful investigation tool
\cite{Aguado:2004,Kiesslich:2006}. In a recent experiment,
fluctuations of the current through two coherently coupled quantum
dots were measured \cite{Barthold:2006}, and the observed,
unexpected super-poissonian noise was subsequently explained as a
pure effect of quantum coherence \cite{Kiesslich:2007}. In the
present work, we consider the same system, however, using an approach based on
the theory of Full Counting Statistics
for systems with Non-Markovian Dynamics (FCSNMD)
\cite{Braggio:2006,Flindt:2007}. The main advantage of this
technique lies in the possibility to calculate all orders of the
current cumulants using a unified approach that allows us to treat
not only the Markovian, but \emph{also} the non-Markovian dynamics
of the charge transport process. We show that the observed,
asymmetric super-poissonian behavior is determined by the interplay
between memory effects induced by the coherent dynamics and the
coupling to the dissipative phonon bath. This indicates that it may
be disputable to associate super-poissonian noise purely to coherent
effects. We conclude that current fluctuations in general can be
strongly modified by memory effects that may be of both intrinsic
and/or extrinsic origin to the system \cite{Flindt:2007}.

We consider the model of transport through a charge qubit in a
dissipative environment previously studied in Refs.\
\cite{Aguado:2004,Kiesslich:2006}. The model
consists of a double quantum dot operated in the strong Coulomb
blockade regime and tuned to a degeneracy point, where
electron transport proceeds
only via the three charge states $|0\rangle$ (unoccupied dots),
$|+\rangle$/$|-\rangle$ (left/right dot occupied). The Hamiltonian
of the system reads
\begin{equation}
\hat{H}=\frac{\varepsilon}{2}\hat{\sigma}_z+T_c\hat{\sigma}_x+
        V_B\hat{\sigma}_z+H_B+\hat{H}_T+\hat{H}_{\mathrm{res}},
\end{equation}
 where we denote the tunnel coupling between $|+\rangle$ and $|-\rangle$ by
$T_c$ and the energy difference (detuning) of the two states by
$\varepsilon$, having introduced the standard spin-boson pseudo-spin
operators $\hat{\sigma}_z\equiv |+\rlangle +|-|-\rlangle -|$
 and $\hat{\sigma}_x\equiv |+\rlangle -|+|-\rlangle +|$. Relaxation
 and dephasing are assumed to occur due to the coupling
$V_B\hat{\sigma}_z=g
\hat{\sigma}_z\sum_j(\hat{a}_j^{\dagger}+\hat{a}_{j})/\sqrt{2}$ to a
surrounding dissipative heat bath described as a reservoir of
non-interacting bosons,
$H_B=\sum_j\hbar\omega_j\hat{a}_j^{\dagger}\hat{a}_j$. We are thus
dealing with the well-known spin-boson problem. Finally, we
introduce a left/right ($\alpha=+,-$) lead described as
non-interacting fermions, \emph{i.e.},
$\hat{H}_{\mathrm{res}}=\sum_{k,\alpha=\pm}\epsilon_{k,\alpha}\hat{c}^{\dagger}_
{k,\alpha}\hat{c}_{k,\alpha}$ coupled to the spin-boson system via
the tunnel-Hamiltonian $\hat{H}_T=\sum_{k,\alpha=\pm}(t_\alpha
\hat{c}^{\dagger}_{k_\alpha}|0\rlangle \alpha|+\mathrm{h.c.})$.

To describe charge transport through the system we follow the
approach developed by Gurvitz and Prager \cite{Gurvitz:1996}.
Assuming a constant tunneling density of states and a large bias
across the double-dot, we can derive an equation of motion for the
reduced density matrix $\hat{\rho}^{(n)}$, defined in the Hilbert
space of the electronic system ($|0\rangle,|+\rangle, |-\rangle$)
and the phonon bath, resolved with respect to the number of
electrons $n$ which have tunneled from the left barrier
\cite{Gurvitz:1996}. To calculate the current cumulants it is
convenient to introduce the \textit{counting field} $\chi$ and
consider the Fourier transform $\hat{\rho}^{(\chi)}=\sum_n e^{i n
\chi }\hat{\rho}^{(n)}$ \cite{Levitov:1994}. Current moments can be
obtained from the dynamics of the sum of the $\chi$-dependent
occupation probabilities contained in the vector
$\mathbf{P}(\chi,t)$ with elements $[\mathbf{P}(\chi,t)]_i=\langle
 i|\mathrm{Tr}_B\{\hat{\rho}^{(\chi)}(t)\}|i\rangle$, $i=0,+,-$.
Here, $\mathrm{Tr}_B$ denotes a partial trace over the bath degrees
of freedom and it is easy to verify the normalization for $\chi=0$.
We proceed by assuming a \textit{local} Born approximation for the
diagonal parts of the density matrix: at any time the bath is
assumed quickly to reach
 local equilibrium corresponding to the given charge state.
Within this approximation we may solve the problem for small values
of $T_c$.\footnote{As we shall see our theory agrees well with a
perturbative approach developed in the hybridized basis which is
exact to all orders in $T_c$ \cite{Aguado:2004}.} The rate equation
for the counting probability vector is
\mbox{$\dot{\mathbf{P}}(\chi,t)=\int_{0}^{t}d\tau\mathbf{W}(\chi,t-\tau)\mathbf{P}(\chi,\tau)$}
with the memory kernel $\mathbf{W}(\chi,t)$ containing the rates for
tunneling to and from the leads and phonon-assisted rates for
tunneling between the two dots. In the following we apply some
recently developed results for the general theory of the FCSNMD to
calculate the current and the noise \cite{Flindt:2007}. The starting
point is the non-Markovian equation of motion for
$\mathbf{P}(\chi,t)$ which in Laplace space translates to
\mbox{$[z-\mathbf{W}(\chi,z)]\mathbf{P}(\chi,z)=\mathbf{P}(\chi,t=0)$}
with $\mathbf{P}(\chi,t=0)$ being the initial condition at
$t=0$.\footnote{For the zero-frequency fluctuations considered
  here, the initial conditions do not play a role.} For the problem at
hand, the memory kernel reads
\begin{equation}
\label{eq:NonMarkGMEZ}
\mathbf{W}(\chi,z)=
\begin{pmatrix}
  -\Gamma_+ & 0 & \Gamma_- e^{i\chi} \\
  \Gamma_+ & -\Gamma^{(+)}_z & \Gamma^{(-)}_z \\
  0 & \Gamma^{(+)}_z & -\Gamma_--\Gamma^{(-)}_z
\end{pmatrix}
\end{equation}
where $\Gamma_\pm=2\pi\varrho_\pm|t_\pm|^2$ are the lead
tunneling rates 
for energy independent densities of states
$\varrho_\pm$. 
The phonon
assisted hopping rates are
$\Gamma^{(\pm)}_z=T^2_c[\mathcal{G}^{(\pm)}(z_+)+\mathcal{G}^{(\mp)}(z_-)]$,
where $z_\pm=z+\Gamma_-/2\pm i\varepsilon$ and the bath correlation
functions in Laplace space are defined as
$\mathcal{G}^{(\pm)}(z)=\!\!=\!\!\int_0^\infty\!\!\!\!dt e^{-zt}
\mathcal{G}^{(\pm)}(t)$. In time domain these functions read
$\mathcal{G}^{(\pm)}(t)=\mathrm{Tr}_B\{ e^{-i H_B^{(+)}
t}\rho^{(\pm)}_\beta e^{i H_B^{(-)}t}\}$ with the equilibrium
density matrix $\rho_\beta^{(\pm)}=e^{-\beta
H_B^{(\pm)}}/\mathrm{Tr}_B\{e^{-\beta H_B^{(\pm)}}\}$ of the
displaced phonon bath with Hamiltonians $H_B^{(\pm)}=H_B\pm V_B$ and
inverse temperature $\beta=1/k_BT$. Assuming an Ohmic spectral
density, $J(\omega)=2 g^2 \omega e^{-\omega/\omega_c}$, we obtain to
order $g^2$ in the bath coupling and to leading order in
$1/\beta\omega_c$, the expression
\begin{equation}
\mathcal{G}^{(\pm)}(z)=\frac{1}{z}
\left\{1-2g^2
    \left[
        H
        \left(
                \frac{\beta z}{2\pi}
        \right)
        +V^{(\pm)}
        \left(
                \frac{z}{\omega_c}
        \right)
    \right]
\right\}
\label{eq_GZ}
\end{equation}
where $H(x)=\ln(x)-\Psi(x)-1/2x$ with $\Psi(x)$ being the digamma
function and $V^{(\pm)}(x)=[\sin(x)\mp i
\cos(x)][\pi/2-\mathrm{si}(x)\mp i \mathrm{ci}(x)]$, where
$\mathrm{si}(x)=\int_0^x dt \sin(t)/t$ and
$\mathrm{ci}(x)=-\int_x^\infty dt \cos(t)/t$. We note that even without
the phonon bath, corresponding to $g=0$, we still have a
contribution of the form $\mathcal{G}^{(\pm)}(z)=1/z$ occurring due
to the coherent dynamics, and only if we take the $z\to 0$ limit we
recover the standard Fermi's golden rule rates for incoherent
tunneling between the dots, $\Gamma^{(\pm)}_{z\to0}=T^2_c
\Gamma_-/[(\Gamma_-/2)^2+\epsilon^2]$.
\begin{figure}[h]
\leavevmode
\includegraphics[width=\linewidth]{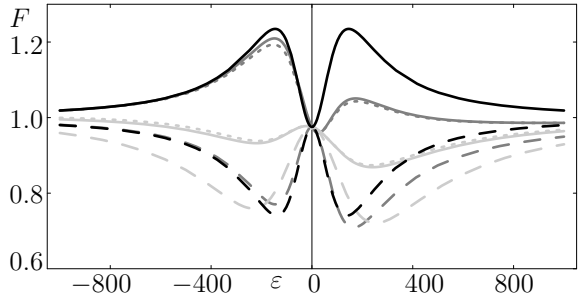}
\caption{Fano factor vs. level detuning $\varepsilon$ for different
physical parameter regimes and two temperatures $T=1.4$ K (light
grey) and $T=12$ K (dark grey). The other parameters correspond to
those used in Fig.\ 2b of Ref.\ \cite{Kiesslich:2007}. Coherent
regime (solid black line), $g=0$ sequential tunneling (black dashed
line), phonon-assisted incoherent tunnelling (dashed grey lines),
phonon-assisted tunneling (solid grey lines) and, for comparison,
perturbative approach in the hybridized basis (dotted lines).}
\label{figure}
\end{figure}
The current $I$ and the noise Fano factor $F=S(0)/eI$ for the system
can be obtained using the theory of FCSNMD
\cite{Braggio:2006,Flindt:2007}.\footnote{The theory also allows us
to calculate higher cumulants, which we, however, will not consider
here.} The theory shows that noise and higher order cumulants, but
not the mean current, may contain signatures of non-Markovian
dynamics, signaled by the presence of $z$-derivatives of the rates
$\Gamma^{(\pm)}_z$. The current is given as $I=\Gamma_- \Gamma_+
\Gamma^{(+)}_0/\Gamma_t^2$ with $\Gamma_t^2=\Gamma_- (\Gamma_+ +
\Gamma^{(+)}_0) + \Gamma_+ \Gamma_p$ and $\Gamma_p=(\Gamma^{(-)}_0 +
\Gamma^{(+)}_0)$. The Fano factor $F$ reads
\begin{equation}
\begin{split}
\label{Fano} &F=F_m+\frac{2\Gamma_-\Gamma_+^2 [(\Gamma_- +
\Gamma^{(-)}_0)\partial_z\Gamma^{(+)}_0-\partial_z\Gamma^{(-)}_0
\Gamma^{(+)}_0]} {\Gamma_t^4},\\
&F_m=\frac{2\Gamma_- \Gamma_+^2 \Gamma^{(-)}_0 + \Gamma_+^2
\Gamma_{p}^2 + \Gamma_-^2 [\Gamma_+^2 +
(\Gamma^{(+)}_0)^2]}{\Gamma_t^4},
\end{split}
\end{equation}
where $F_m$ denotes the Markovian contribution to the Fano factor
(taking only the $z=0$ contribution) and where we have defined
$\partial_z\Gamma^{(\pm)}_0\equiv\partial_z[\Gamma^{(\pm)}_z]_{z=0}$.

In Fig.\ \ref{figure} we show the Fano factor calculated in
different parameter regimes. Without coupling to the phonons
($g=0$), we find in the limit  $z\to 0$ the \textit{sequential
tunneling} regime described only by the Markovian terms
\cite{Braggio:2006}. The Fano factor is symmetric around
$\epsilon=0$ and sub-poissonian (black dashed curve). Taking the
full $z$-dependence of the rates, non-Markovian contributions are
included, and we find the so-called \textit{coherent} regime
\cite{Gurvitz:1996,Nazarov:1996}. The Fano factor is again
symmetric, but now super-poissonian (solid black curve)
\cite{Kiesslich:2007}. Including coupling to the phonon bath
($g>0$), we are in the \textit{phonon-assisted} regime, where
electron tunneling between the dots is assisted by phonon absorption
and/or emission. In the \textit{incoherent} approximation without
non-Markovian contributions, we find sub-poissonian noise
independently of the temperature (dark and light grey dashed lines).
On the other hand, including non-Markovian terms we see the
asymmetric super-poissonian behavior for low temperatures (dark grey
solid lines) which disappears with increasing temperature (light
grey solid lines) as also seen in the experiment
\cite{Barthold:2006}. We, moreover, find an optimal agreement with
the results of the standard perturbation theory in the hybridized
basis (dark and light grey dotted lines), allowing us to exclude
possible spurious effect of the perturbation in $T_c$. We see that
the noise becomes super-poissonian due to the non-Markovian
corrections. By direct inspection of the expression for the Fano
factor, we see that the condition for having super-poissonian noise
is given by the inequality $\Gamma_+(\partial_z\Gamma^{(+)}_0
(\Gamma_- +\Gamma^{(-)}_0)-\Gamma^{(+)}_0
\partial_z\Gamma^{(-)}_0) > \Gamma^{(+)}_0
(\Gamma_p+\Gamma_-+\Gamma_+)$. This shows that a pure coherent
contribution cannot easily be disentangled from other classical or
quantum bath memory effects, since these all appear mixed in the
 memory kernel. Our analysis supports the suspicion,
previously arisen, that an anomalous behavior of the current
cumulants may be due to classical memory effects \cite{Flindt:2007}.
Such effects, independently of their origin, can destroy the
signatures of pure quantum effects, and should be carefully taken
into account in any interpretation of current fluctuations.

Support by the European Science Foundation
(ESF) Research Networking Programme entitled ``Arrays of Quantum Dots and
Josephson Junctions'' and by EU via MCRTN-CT2003-504574 network and by the grant 202/07/J051 of the Czech Science Foundation are gratefully acknowledged.

\end{document}